\documentstyle[multicol,epsfig,aps,prb,array]{revtex}
\newcommand \be{\begin{equation}}
\newcommand \ee{\end{equation}}
\newcommand \ba{\begin{eqnarray}}
\newcommand \ea{\end{eqnarray}}
\begin{document}
\title
{\Large\bf
Lateral current density fronts in asymmetric
double-barrier resonant-tunneling
structures}

\author{\bf
Pavel Rodin \cite{LEAVE}
and
Eckehard Sch{\"o}ll}
\address
{Institut f{\"u}r Theoretische
Physik, Technische Universit{\"a}t Berlin,\\
Hardenbergstrasse 36, D-10623, Berlin, Germany
}
\setcounter{page}{1}
\date{\today}
\maketitle

\hyphenation{cha-rac-te-ris-tics}
\hyphenation{se-mi-con-duc-tor}
\hyphenation{fluc-tua-tion}
\hyphenation{fi-la-men-ta-tion}
\hyphenation{self--con-sis-tent}
\hyphenation{cor-res-pon-ding}
\hyphenation{con-duc-ti-vi-ti-tes}


\abstract
We present a theoretical analysis and numerical simulations
of lateral current density fronts in bistable resonant-tunneling diodes with
Z-shaped current-voltage characteristics.
The bistability is due to the charge accumulation in
the quantum well of the double-barrier structure.
We focus on asymmetric structures in the regime
of sequential incoherent tunneling and study the dependence
of the bistability range,
the front velocity and the front width on the structure parameters.
We propose a sectional design of a structure that is
suitable  for experimental observation of front propagation and discuss
potential problems of such measurements in view of our theoretical
findings.  We point out the possibility to use sectional resonant-tunneling
structures as controllable three-terminal switches.

\endabstract

\vspace{.5cm}

PACS {05.70.Ln,72.20.Ht,85.30.-z}

\begin{multicols}{2}

\newpage


\section{Introduction}

Charge accumulation in the well of a double-barrier resonant tunneling (DBRT)
structure leads to electrostatic feedback that increases the energy
of the quasibound state with respect to the
emitter, thus making resonant tunneling possible for higher applied
voltages ~\cite{GOL87}. This may result in bistability when the high current
density and low current density states coexist in a certain range of
the applied voltage. The current-voltage characteristic becomes Z-shaped instead
of the conventional N-shape  ~\cite{GOL87,MAR94}. It has been pointed out that
the bistability provides a basis for formation of lateral
current density patterns like traveling current density fronts
~\cite{WAC95d,GLA97,MEL98,FEI98}.
Once triggered, such a front leads to switching of the structure between the
on-- and off--state,
depending on the applied voltage ~\cite{GLA97,MEL98,FEI98,MEI00}.
Front propagation in the DBRT provides a unique example of nonlinear
pattern formation in a system with
quantum transport. In the context of the DBRT studies,
it represents a novel switching process ({\em trigger front})
which is spatially inhomogeneous and is likely to determine
the general properties of switching dynamics in large-area DBRT
structures.

Recent theoretical work ~\cite{GLA97,MEL98,FEI98,MEI00,CHE00,SCH02} has
essentially focused on the {\it qualitative} behavior of current density fronts,
whereas the possibility of their experimental observation crucially depends
on {\it quantitative} parameters, first of all on the front propagation
velocity and the  front width. This is due to the limits of temporal resolution
and finite lateral dimensions of DBRT structures.
In this article we aim to bridge this gap
and present the results of numerical simulations of front dynamics
in asymmetric DBRT structures that provide a wide range of bistability
and thus are favorable for experimental observation.
We also discuss the specific design of DBRT structures
that are suitable for such experiments.

\section{Model of the DBRT structure}

The bistability of a DBRT structure occurs due to the charge accumulation in the
quantum
well ~\cite{GOL87,MAR94}. As a result of the electrostatic feedback
the potential of the bottom of the
well $\Phi$  depends not only on the applied voltage $u$,
but also on the  built-up electron concentration  $n$  (Fig.1).
For structures with narrow well and thick barriers
the $\Phi(u,n)$ dependence can be calculated by
representing electrons in the well as a sheet charge of negligible
width ~\cite{CHE00}:
\begin{equation}
\Phi(u,n) = - \frac{b_1}{d} u + \frac{b_1 b_2}{d}
\frac{e n (x,y)}{\epsilon \epsilon_0},
\qquad
d = b_1 + b_2.
\label{bottom_well}
\end{equation}
Here $u < 0$ is the applied voltage, $e < 0$ is the electron charge,
$\epsilon$ and $\epsilon_0$ are the relative and absolute
permittivities of the material, respectively. The effective widths
of the emitter-well barrier $b_1$ and the well-collector barrier
$b_2$ include the half-width of the well,
and $b_2$ also includes the width of the
depleted collector layer (Fig.1)
~\cite{COMMENT_0}. Eqn.(\ref{bottom_well}) is
also applicable in the case when the built-up concentration
is inhomogeneously distributed in the transverse $(x,y)-$plane, provided that
the transverse variations of $n(x,y,t)$ are smooth in the
sense that their characteristic length is much larger than the
effective thickness of the structure $d$.
Below we will show that this condition is indeed met for current
density fronts.

In the regime of sequential tunneling the DBRT structure can be
described by the  continuity equation for the electron density
per unit area in the well $n(x,y,t)$ ~\cite{WAC95d,GLA97,FEI98,MEI00,SCH02}
\begin{eqnarray}
\label{CONT}
\frac{\partial n}{\partial t} + \frac{1}{e}\nabla_{\perp} {\bf
J}_{\perp} &=&
\frac{1}{e}\left(J_{ew} - J_{wc} \right), \\
\nonumber
\nabla_{\perp}
& \equiv &
{\bf e}_x\frac{\partial }{\partial x}+{\bf
e}_y\frac{\partial}{\partial y},
\end{eqnarray}
where $J_{ew}(x,y)$ and $J_{wc}(x,y)$ are local densities (per unit area)
of the emitter-well and the well-collector currents, respectively,
${\bf J}_{\perp}$ is the density (per unit length) of the transverse current
in the well.

The transverse redistribution of charge in the well is treated
semiclassically in the drift-diffusion approximation according to
Ref.\onlinecite{CHE00}.
The drift component of ${\bf J}_{\perp}$
is proportional to the lateral electrical field
${\bf {\cal E}}_{\perp} = -{\bf \nabla}_{\perp} \Phi$. Eq.(\ref{bottom_well})
yields ${\bf {\cal E}}_{\perp} \sim {\bf \nabla}_{\perp} n$, and hence the drift
current,
similar to the diffusive current,
turns out to be proportional to the gradient of the built-up electron
concentration.
This allow us to represent the transverse current density in the well
in terms of an effective concentration-dependent diffusion coefficient
$D(n)$~\cite{CHE00}:
\begin{eqnarray}
\label{transverse_transport}
{\bf J}_{\perp} &=& - e D(n) \nabla_{\perp} n,  \\
\nonumber
D(n) &=& \frac{|e| \mu n}{4 \epsilon \epsilon_0}
\left(
\frac{4 b_1 b_2}{d} +
\frac{r_B}{1 - \exp[- n / (\rho_0 k T_W)]}
\right).
\end{eqnarray}
Here $\mu$ is electron mobility in the well,
$r_B =(4 \pi \epsilon \epsilon_0 \hbar^2)/(e^2 m)$ is Bohr radius in the
semiconductor material, $\rho_o = m/(\pi \hbar^2)$ is the 2D density
of states, $m$ is the effective electron mass and $k$ is Boltzmann's
constant. The derivation of
Eq.(\ref{transverse_transport}) assumes a
Fermi distribution in the well ~\cite{CHE00}.
The electron temperature $T_W$ in the well can be
different from the substrate temperature $T$.
In the following we assume a substrate temperature $T = 4$~K.

The emitter-well and well-collector current densities
are modelled by simple approximate formulas
~\cite{MEI00,SCH02,COMMENT_1}
\begin{eqnarray}
J_{ew} &=& \frac{e \Gamma_L}{\hbar} \left[\rho_o \Delta - n \right]
\nonumber \\
&\times&
\frac{\arctan \left(2\Delta/\Gamma_W \right)
- \arctan \left( 2 \Omega / \Gamma_W \right)}{\pi},
\label{current_density_1}\\
J_{wc} &=& \frac{e \Gamma_R}{\hbar} n,
\label{current_density_2}
\end{eqnarray}
where
\begin{equation}
\Delta(n,u) \equiv E_e^F - E_W - e\Phi, \;
\Omega(n,u) \equiv - E_W - e\Phi
\end{equation}
denote the energy of the quasibound state with respect to the Fermi
level and the bottom of the conduction band in the emitter,
respectively, $E_e^F$ is the Fermi energy in the emitter,
$E_W > E_e^F$ is the energy of the quasibound state
with respect to the bottom of the well, $\Gamma_{L}$
and $\Gamma_{R}$
are the energy broadenings associated with the emitter-well
and the well-collector barriers, respectively,
$\Gamma_W = \Gamma_L + \Gamma_R + \Gamma_{scatt}$ is the total
broadening of the quasi-bound state resulting from the escape
{\it via} emitter-well barrier, well-collector barrier and
scattering in the well, respectively. The broadening of the
quasi-bound state is characterized
by a Lorentzian spectral function.
The assumption of  sequential tunneling
implies that $\Gamma_L, \Gamma_R < \Gamma_{scatt}$.
Similar models for vertical transport
have also been used in Refs.\onlinecite{FEI98,INK01}.

Eqn.(\ref{current_density_1}) has a transparent physical meaning:
$\Gamma_L / \hbar$ is the tunneling rate between the emitter and the well;
$ [\rho_0 \Delta - n]$ is the difference between the number of transverse
tunneling modes available for the given position of the quasibound state
~\cite{DAT95} and the number of occupied states in the well;
the last factor accounts for smooth decrease of current
near the resonance breaking points
$\Delta = 0$ and $\Omega = 0$. Eqn.(\ref{current_density_2})
describes the one-way tunneling from the well to the collector with
a rate $\Gamma_R / \hbar$. In the stationary state the current
density $J$ is proportional to the electron concentration $n$ stored
in the well. Formal derivation of this transport
model ~\cite{MEI00,SCH02}
is based on  the following essential assumptions:

(i) barriers are high in the sense that the matrix
element of the emitter-well and well-collector transitions do
not depend on the vertical (in $z-$direction) electron momentum,
hence the barriers are fully characterized by constant
transparencies $\Gamma_{L,R}$;

(ii) the temperature of the substrate $T$
and the electron temperature of the well $T_W$ are small compared to
$E^F_e$ and  $\Gamma_W$  ;

(iii) the Fermi level in the collector is lower than the bottom of the
well.

Combining (\ref{bottom_well})--(\ref{current_density_2})
and taking into account only one transverse dimension,
we obtain the following nonlinear parabolic
equation for the built-up electron charge
\begin{eqnarray}
\label{CONT_1}
\frac{\partial n}{\partial t} &=&
\frac{\partial}{\partial x} \left( D(n) \frac{\partial n}{\partial x} \right)+
f(n,u),
\\
\nonumber
f(n,u) &=& \frac{1}{e} \left(J_{ew}(u,n) - J_{wc}(n) \right).
\end{eqnarray}
In the stationary state $J = J_{ew} = J_{wc}$, and the local current--voltage
characteristic $J(u) = J(u,n(u))$ is determined by
the stationary $n(u)-$dependence together with
Eqns.(\ref{current_density_1}),(\ref{current_density_2}).
The $n(u)-$dependence, in turn, follows from the steady-state
condition $f(n,u)=0$. The $J(u)$ characteristic is bistable for
sufficiently small broadening of the quasi-bound state $\Gamma_W$.
A typical example of such Z-shaped characteristics is shown in Fig.2.

We use the following set of material and structure parameters:
$\epsilon = 12$ and  $m = 0.067 \, m_0$ (for GaAs),
where $m_0$ is the free electron mass;
$E_e^F = 10 \; {\rm meV}$, $\mu = 10^5 \; {\rm cm^2/ Vs}$,
$E_W = 40 \; {\rm meV}$, $T = 4 K$. The effective width of the
emitter-well barrier is taken as $b_1 = 10 \; {\rm nm}$.
The effective width of the second barrier $b_2$
includes the width of the depletion
layer in the collector
$l_d \sim (\epsilon \epsilon_0 u)/(e N_D d)$ (see Fig.1),
which lies in the interval
$10...100 \; {\rm nm}$ for the typical doping level
$N_D = 10^{17} \; {\rm cm^{-3}}$ in the collector ~\cite{COMMENT_0}.

The dependence of the transparencies $\Gamma_{L}$ and $\Gamma_{R}$
on the actual barrier widths  can be evaluated by solving the
one-dimensional Schr{\"o}dinger equation
~\cite{PRI92}. Compact formulas for the transparencies of
GaAs/AlGaAs barriers are given in Ref.\onlinecite{GUE89}.
Typical values are $\Gamma_{L,R} \sim 0.1...1~$meV  for barrier
widths of the order of 10~nm ~\cite{GUE89}.
The corresponding tunneling time is
$\tau_T \sim 0.06...0.6~{\rm ps}$.
The parameter $\Gamma_{scatt}$  can be estimated on the basis
of the electron mobility in the quantum well.
Typical mobility values
$\mu \sim 10^4...10^5~{\rm cm^2/Vs}$ yield
$\Gamma_{scatt} \sim e \hbar / m \mu  \sim 0.1 ... 1~ {\rm meV}$
~\cite{RID95,KAR90}.
The corresponding momentum relaxation time
$\tau_m$ is of the order of 1~ps or less.
Below we assume the total broadening of the quasibound state to be
$\Gamma_W = 1 ... 2~{\rm meV}$, and take into account that
$\Gamma_L, \Gamma_R < \Gamma_W$.
After tunneling from the emitter to the quantum well
the transverse kinetic energy of an electron
is transferred to the lattice.
The temperature of the electron gas in the well
depends on the relation between the energy relaxation time $\tau_e$
and the tunneling time $\tau_{T}$ ~\cite{times}.
The limit cases $\tau_{e} \ll \tau_{T}$ and
$\tau_{e} \gg \tau_{T}$ correspond to a
cold ($T_W \approx T$) and hot ($T_W > T$) electron gas, respectively.
Since for resonant tunneling conditions the transverse kinetic energy is limited
by $E^F_e$,
the upper bound for the electron
temperature is given by  $T_W = E^F_e / k \approx 100$~K.
Below we consider the whole range of relevant temperatures
$T < T_W < E^F_e / k$
and find that the effect of electron
heating on the front dynamics is negligible.

\section{Range of bistability}

We shall start with two simple analytical estimates for the
range of bistability $\Delta u \equiv |u_{th} - u_{h}|$,
where $u_{h}$ and $u_{th}$ are the left and right turning points
of the Z-shaped current-voltage characteristic that correspond
to the holding and threshold voltages, respectively (see Fig.2).
In the limit $\Gamma_W \ll E_e^F$
the current--voltage characteristic becomes piecewise linear.
Then the ultimate upper bound for the bistability range $\Delta u$
immediately follows from
Eqns.(\ref{current_density_1}),(\ref{current_density_2}) and the
condition $J_{ew}=J_{wc}$  (see also (\ref{uthuh}) in the Appendix):
\begin{equation}
\Delta u =
\frac{1}{|e|}
\frac{4 \Gamma_L}{\Gamma_L + \Gamma_R} \frac{b_2}{r_B} E_e^F.
\label{bistability_range_1}
\end{equation}
Eq.(\ref{bistability_range_1}) indicates
that low transparency $\Gamma_R$ and large effective thickness $b_2$
of the well-collector barrier are favorable for bistability.
(Note that with increasing $d$ the intrinsic capacitance of the
well becomes smaller, enhancing the electrostatic feedback from
the built-up electron charge.)

For finite broadening of the quasibound state $\Gamma_W$
the bistability range shrinks or even disappears.
This effect can be evaluated analytically by using
a square-shaped spectral function
\begin{eqnarray}
\nonumber
A_W(E,{\bf k}) &=& \frac{1}{\Gamma_W}
( \Theta(E-E_k-E_W-e\Phi_W+\Gamma_W/2)
\\
&-& \Theta(E-E_k-E_W-e\Phi_W-\Gamma_W/2)),
\label{spectral_2}
\end{eqnarray}
where $\Theta$ is the Heaviside function,
instead of the Lorentzian spectral function
\begin{equation}
A_W(E,{\bf k}) = \frac{\Gamma_W}{\left[(E-E_k-E_W-e\Phi_W)^2+\Gamma_W^2/4
\right]},
\label{spectral_1}
\end{equation}
that has been used to
derive Eqns.(\ref{current_density_1}),(\ref{current_density_2}).
(Here $\bf{k}$ is the transverse electron momentum and $E_k$ is the
corresponding kinetic energy.)
This substitution yields
\begin{equation}
\Delta u \approx
\frac{1}{|e|}
\left[
\frac{4 \Gamma_L}{\Gamma_L + \Gamma_R} \frac{b_2}{r_B}
\left( E_e^F - \frac{\Gamma_W}{2} \right) - \frac{d}{b_1} \Gamma_W
\right].
\label{bistability_range_2}
\end{equation}
According to (\ref{bistability_range_2}),
the effect of geometrical asymmetry $b_1 \neq b_2$ is not
monotonic: for fixed total width $d = {\rm const}$ there is an optimal
ratio
\begin{equation}
\frac{b_1}{d} =
\sqrt{\frac{\Gamma_L + \Gamma_R}{2\Gamma_L}
\frac{r_B}{d} \frac{ \Gamma_W}{2 E_e^F - \Gamma_W}}.
\end{equation}

Eqn.(\ref{bistability_range_2}), along with
Eqn.(\ref{bistability_range_1}),
predicts that  $\Delta u$ monotonically
increases with decrease of the $\Gamma_R/\Gamma_L$ ratio.
In contrast to this conclusion,
the ranges of bistability calculated numerically for the full
model (\ref{current_density_1}),(\ref{current_density_2}) (see Fig.3)
show that there is an optimal ratio  $\Gamma_R/\Gamma_L$,
roughly  $\Gamma_R/\Gamma_L \approx 0.2$, that corresponds to the
widest bistability range.
When $\Gamma_R$ is too small,
the bistability vanishes due to the relative increase of
current density in the off-state.
Eqn.(\ref{bistability_range_2}) does not catch this effect
because the oversimplified spectral function (\ref{spectral_2})
inadequately describes the low-branch of the current--voltage characteristic
~\cite{COMMENT_GLA}.

\section{Stationary propagating fronts}

{\it General properties.}
Stationary moving fronts correspond
to self-similar solutions of Eqn.(\ref{CONT_1})
$n(x,t) = n(x-v t)$ satisfying the
boundary conditions $n(- \infty) = n_{on}$,
$n(+\infty) = n_{off}$, where
$n_{on}$ and $n_{off}$ are the electron concentrations in
the well in the on- and off-states, respectively.
Such solutions give a good approximation for
fronts in a finite system with lateral size $L$
much larger than the front width $W$, regardless
of the actual boundary conditions for built-up
concentrations at the lateral edges of the DBRT. A positive velocity $v > 0$
corresponds to propagation of the on-state into the off-state (hot front),
a negative velocity corresponds to propagation of the off-state
into the on-state (cold front); this triggers switching between the two
states. In the co-moving frame
$\xi = x - v t$ Eqn.(\ref{CONT_1})
becomes an ordinary differential equation
\begin{equation}
\frac{d}{d \xi} \left( D(n) \frac{d n}{d \xi} \right) +
v \frac{d n}{d \xi} + f(n,u) = 0.
\label{comoving}
\end{equation}
Multiplying by
$D(n) dn / d \xi $ and integrating over
$\int_{-\infty}^{+\infty} d\xi$ yields \cite{SCH00}
\begin{equation}
v = \frac{ \int_{n_{off}}^{n_{on}} f(n,u) D(n) dn }
{\int_{-\infty}^{\infty} (dn/d \xi)^2 D(n) d \xi}.
\label{velocity}
\end{equation}
The direction of front propagation is determined by the sign of
the numerator in Eq.(\ref{velocity}), and hence by the applied voltage $u$.
The equal areas rule $\int_{n_{off}}^{n_{on}} f(n,u_{co}) D(n) dn$
corresponds to $v=0$ and specifies the voltage
$u_{co}$ at which a stationary current density
front exists.

{\it Formula for the front velocity.} In the limit case of small broadening of
the quasibound state $\Gamma_W \ll E^F_e$
equation (\ref{comoving}) can be integrated analytically,
if we additionally assume $D(n) = D_0 = const$ (see Appendix).
This leads to the following approximate $v(u)-$dependence
\begin{equation}
v = \sqrt{\frac{\Gamma_L D_0}{\hbar}}
\frac{ (1 + \sqrt{\alpha_{RL}} A)^2}{2 A}
\frac{r_B}{b_2}
\frac{|e| (u - u_{co})}{E^F_e},
\label{v(u)}
\end{equation}
where
$$
\lambda \equiv \frac{4 b_1 b_2}{d r_B},
\;
\alpha_{RL} \equiv \frac{\Gamma_R}{\Gamma_L},
\;
A \equiv \sqrt{\lambda + \alpha_{RL} + 1} + \sqrt{\alpha_{RL}}
$$
and $u_{co}$ is determined  by (\ref{u_co}). Generally,
Eqn.(\ref{v(u)}) is applicable to slow fronts
when $|e(u - u_{co})| \ll E^F_e$.

The diffusion coefficient $D_0$ can be estimated
on the basis of Eqn.(\ref{transverse_transport}),
assuming drift-dominated transport in the well and choosing
an average concentration $ \left< n \right>$ in the front wall
\begin{eqnarray}
\label{diffusion}
D_0 &\approx& \frac{|e| \mu b_1 b_2}{ \epsilon \epsilon_0 d}
\left< n \right> = \frac{\mu \lambda}{|e| \rho_o} \left< n \right>,
\\  \nonumber
\left< n \right> &=& \frac{n_{on}(u_{co}) + n_{off}(u_{co})}{2},
\end{eqnarray}
where $n_{on}(u_{co})$ and $n_{off}(u_{co})$
are the electron concentrations in the well for
on- and off-states at $u = u_{co}$, respectively.
(We note that the assumption of drift-dominated transport
is supported by the numerical results presented below.)
Using (\ref{concentrations}) and (\ref{u_co}),
we substitute (\ref{diffusion}) into (\ref{v(u)}) and get
\begin{eqnarray}
v &=& \sqrt{\frac{\mu \Gamma_L E^F_e }{e \hbar}} \\
\nonumber &\times&
\sqrt{\frac{(1 + \sqrt{\alpha_{RL}} A)^3(1 + \lambda +
\sqrt{\alpha_{RL}} A)}
{\lambda (\lambda + \alpha_{RL} + 1) A^2}} \\
\nonumber &\times&
\frac{b_1}{d}
\frac{|e| (u - u_{co})}{E^F_e}.
\label{v(u)2}
\end{eqnarray}
According to (\ref{bistability_range_1})
the width of the bistability range $\Delta u$
is of order of $E^F_e$,
and the second factor in (\ref{v(u)2}) is of the order of
unity for realistic structure parameters. Hence the order of
magnitude value of the front velocity is given by
\begin{equation}
v \sim
\sqrt{\frac{\mu \Gamma_L E_e^F }{e \hbar}}.
\label{v_max}
\end{equation}
We get $v \sim 10^7 \, {\rm cm/s}$ for $E^F_e = 10 \; {\rm meV}$,
$\Gamma_L = 1 \; {\rm meV}$ and $\mu = 10^5 \; {\rm cm^2/Vs}$.

{\it Numerical results.} Numerical results for the front velocity $v(u)$ and
front
width $W(u)$ are summarized in Fig.4. It follows from
Eqn.(\ref{CONT_1}) that these quantities
scale as
\begin{equation}
v \sim \sqrt{Df}  \qquad {\rm and} \qquad  W \sim \sqrt{\frac{D}{f}},
\label{scaling}
\end{equation}
respectively. The scaling rules (\ref{scaling})
allow one to extend the numerical results to any value
of $\mu$ and $\Gamma_{L,R}$ according to
$v \sim \sqrt{\mu \Gamma_L}$ and $W \sim \sqrt{\mu / \Gamma_L}$,
provided that the ratio $\Gamma_R/\Gamma_L$ and $\Gamma_W$
are kept the same.

The order of magnitude values of the front velocity and
the front widths are  $10^7~{\rm cm/s}$ (in accordance
with (\ref{v_max})) and $10~{\rm \mu m}$,
respectively (Fig.4).
For lateral dimensions of the order
of $100 \; {\rm \mu m}$, the switching times are thus less than 1 ns.
The front width exceeds
both the thickness of the structure $d \sim 50 \, {\rm nm}$
and the mean free path of the electron in the well which
can be estimated as $\ell_m \sim 0.1...1 \; {\rm \mu m}$
for $\mu = 10^5 \; {\rm V/cm s}$ ~\cite{CHE00}.
Hence the local approximation in (\ref{bottom_well})
and the drift-diffusion approximation (\ref{transverse_transport})
for the lateral transport in the well are indeed justified
for the obtained front solutions.
Cold fronts are faster and wider than hot fronts. This is the
result of the higher contrast between $n_{on}$ and $n_{off}$
near the right turning point
$u_{th}$ of the current-voltage characteristic.
The estimate of Eqn. (\ref{v(u)2}) for the same set of
parameters is shown in Fig.4(a) by the dashed line.
Despite of the rough approximations adopted  to derive (\ref{v(u)2})
(see Appendix), it gives the correct order of the magnitude
of the front velocity.

In order to evaluate the significance of  heating of the electron gas
in the well we have performed simulations for $T_W ~\sim E_e^F/k =
100 \; {\rm K}$,
which corresponds to the upper bound of the electron temperature $T_W$
(Fig.4a, curve 2). It is found that  heating  has a visible but minor effect
on front velocity. This indicates the
drift-dominated lateral transport in the quantum well.
To confirm this conclusion we have also performed simulations for
$T_W = 0$, when lateral diffusion in the well is minimized
and have found that the front velocity is practically the same as for
$T_W = T = 4 \, {\rm K}$ (Fig.4a, curve 1).
Generally, according to (\ref{transverse_transport})
the drift component of the lateral current is much larger than the
diffusion component for  $n  \gg \rho_0 kT_W$ ~\cite{CHE00}.
As it follows from the characteristic shown in Fig.2,
for $T = T_W$ this condition is satisfied
for almost the whole range of concentrations $[n_{on},n_{off}]$
within the propagating front.
Note that since the effective
diffusion coefficient in (\ref{transverse_transport}) is
concentration-dependent,
the voltage $u_{co}$ shifts with electron temperature $T_W$.
The right and left extreme
points for $u_{co}$ correspond to the limits $T_W \rightarrow 0$ (pure drift)
and $T_W \rightarrow \infty$ (pure diffusion), respectively.

\section{Discussion: towards experimental observation of front
dynamics.}

Since the pioneering work ~\cite{GOL87} the experimental observation
of bistability in DBRT structures has become a routine. Despite of this, to
the best of our knowledge lateral current density patterns in the DBRT,
predicted already several years ago, have not yet been observed
experimentally. Below we discuss the potential
problems of such measurements in view of our theoretical findings.

{\it Sectional design of DBRT structure.}
Achievements of modern growth technology have
made it possible to manufacture resonant-tunneling structures
with sophisticated design ~\cite{WER99,WER00}, opening  up the way
to study transverse current density patterns in the DBRT experimentally.
Here we propose the simplest design of a DBRT diode with a
sectional emitter that allows
to trigger a current density front and to
observe its propagation by means of electrical measurements.
Such a device constitutes of identical sections with
individual emitters but common collector so that the bias
can be applied separately to each section.
To trigger a hot front the system should be prepared in the uniform
off-state. This can be done by applying a voltage
$u > u_{h}$ to all emitter sections
and subsequently decreasing this voltage to a value $u_h < u < u_{co}$.
Then the DBRT can be switched on locally by a short negative voltage pulse
$u < u_{h}$, applied to either lateral or central section of the
device. The resulting front propagation can be monitored
by measuring the electrical current in the parallel sections of the
device.

{\it Barriers asymmetry.} Our results suggest that to achieve
the widest bistability range  the transparency of the well-collector
barrier should be approximately 5 times smaller than the transparency
of the emitter-well barrier.

{\it Front velocity and time resolution.}
Our simulations show that the order of magnitude
values of the  front velocity and width are $10^7 \, {\rm cm/s}$
and $10 \, {\rm \mu m}$, respectively.
It implies that  the lateral dimension
of suitable  DBRT structure should exceed $100 \; {\rm \mu m}$.
Since it takes about 1 ns for the front  to cross such a structure,
reliable measurements of front propagation demand for
time resolution of 100~ps or better.
The fronts become slower with  decreasing
transparency of the barriers $\Gamma_{L}$ and $\Gamma_{R}$.
However, then the front width $W$ will increase (see Eq.(\ref{scaling})),
setting severe additional  limitations on the lateral
dimension of the appropriate DBRT structure.
For example, decreasing $\Gamma_{L,R}$ by one order of magnitude
to the values
$\Gamma_L = 0.05 \, {\rm meV}$, $\Gamma_R = 0.01 \, {\rm meV}$
results in a front width of the order of $\sim 30 \, {\rm \mu m}$,
whereas the width of the cold front can exceed $100 \, {\rm \mu m}$.
Finally we note that in the regime of coherent
tunneling the front velocity is expected to be of the order of the Fermi
velocity  $v_f = \sqrt{2 E^F_e/m^{\star}}$ ~\cite{GLA97} . This yields
$\sim 10^8 \, {\rm cm/s}$ for $E^F_e = 10 \, {\rm meV}$.

{\it Hot or cold fronts ?}
The width of a hot front is considerably smaller than the width
of  cold fronts and its velocity is somewhat lower (Fig.4).
Hence a hot front is expected to be an easier target for
experimental observation than a cold front.

{\it Pinning of the front.} Pinning on small
imperfections of the reference media has been recognized as
an important mechanism capable of preventing the front
propagation in nonlinear systems ~\cite{MIT98}.
In the case under consideration both
technological imperfections of the DBRT structure
and  the spatial inhomogeneity due to the sectional
DBRT design may play a role. Due to the exponential
dependence of the barrier transparencies $\Gamma_{L,R}$
on the barrier width,  the  embedded fluctuations of the barrier
width are the most important among technological imperfections.
In presence of imperfections the stationary current--voltage
characteristic includes additional stable branches that
correspond to nonuniform current density profiles associated with
imperfections ~\cite{IMP}. These steady non-uniform states remain observable,
though the uniformly propagating fronts may not be possible
in low-quality structures.

{\it Effect of the external circuit.} Generally, bistable
DBRT elements can be operated via an external load.
In contrast to the well known case
of S-type bistability, in the system with Z-type bistability
an external load provides a positive, not negative feedback
upon front dynamics. In particular, stationary current
density patterns are not stable even under current-controlled conditions
~\cite{MEI00}. An active external circuit which simulates
a negative external load resistance and reverses the sign of feedback
has been implemented experimentally in Ref.\onlinecite{MAR94}.
We refer to Refs.\onlinecite{MEI00,ALE98}
for the further discussion of the effect of external circuit on
dynamics and stability of current density patterns.

{\it DBRT-based 3-electrode switch.}
The sectional DBRT with two small (control) sections separated
by a large (main) section can be potentially used in applications
as a controllable switch. By applying a control voltage pulse
to the side sections, the rest of the structure can be switched
to the on-- or off--state via propagation of a lateral current density
front. Note that in semiconductor device physics propagation of
switching fronts in three-electrode bistable devices
is a well known process which is used, e.g, in thyristors in certain regimes
of operation  ~\cite{ZEE,MEI98}.\\

\acknowledgements

We are grateful to A. Wacker, L.E. Wernersson and V. Cheianov
for useful discussions. This work has been supported by DFG in the framework
of Sfb 555. One of the authors (P.R.) acknowledges support
by the Alexander von Humboldt Foundation.

\appendix
\section*{Analytical approximation for the front velocity.}

In the limit $\Gamma_W \ll E_F^e$ the local kinetic function
becomes piecewise linear:
\begin{eqnarray}
\label{fnu}
&&f(n,u) = \frac{e \Gamma_L}{\hbar}(\lambda + \alpha_{RL} + 1)
\left[\frac{\rho_0(e \gamma u - E_W + E^F_e)}
{\lambda + \alpha_{RL} + 1} - n \right] \nonumber\\
& &{\rm for } \qquad
n \in \left[ \frac{\rho_0 (e \gamma u - E_W)}{\lambda};
\frac{\rho_0 (e \gamma u - E_W + E^F_e)}{\lambda} \right],
\nonumber \\ & &
f(n,u) = - \frac{e \Gamma_L}{\hbar} \alpha_{RL} n\\
& &{\rm for} \qquad
n \notin \left[ \frac{\rho_0 (e \gamma u - E_W)}{\lambda};
\frac{\rho_0 (e \gamma u - E_W + E^F_e)}{\lambda} \right],
\nonumber
\end{eqnarray}
where the following notations have been introduced
\begin{equation}
\lambda \equiv \frac{4 b_1 b_2}{d r_B},
\qquad \gamma \equiv \frac{b_1}{d},
\qquad
\alpha_{RL} \equiv \frac{\Gamma_R}{\Gamma_L}.
\end{equation}
For a voltage $u$ within the bistability range
(note that $u < 0$ and $e < 0$ )
\begin{equation}
u_{th} =
\frac{E_W}{e \gamma} + \frac{1}{e \gamma} \frac{\lambda E^F_e}{1 +
\alpha_{RL}}
< u <
u_{h} = \frac{E_W}{e \gamma}
\label{uthuh}
\end{equation}
the electron concentration in the well is given by
\begin{eqnarray}
\label{concentrations}
n_{on} &=& \frac{\rho_0(e \gamma u - E_W + E^F_e)}
{\lambda + \alpha_{RL} + 1};
\\ \nonumber
n_{int} &=& \frac{\rho_0(e \gamma u - E_W)}{\lambda};
\\ \nonumber
n_{off} &=& 0
\end{eqnarray}
for on, intermediate and off states, respectively. The corresponding
current-voltage characteristic is also piecewise linear.

Using (\ref{fnu}) and assuming for simplicity $D(n) = D_0 = const$,
we find an explicit solution of Eqn.(\ref{comoving})
\begin{eqnarray}
n(\xi) &=& n_{int} \exp (\beta_1(v) \xi)
\\ \nonumber
&{\rm for}&
\qquad
0 < n < n_{int} \; {\rm and} \;  - \infty < \xi < 0;
\\ \nonumber
n(\xi) &=& n_{on} - (n_{on} - n_{int}) \exp (-\beta_2(v) \xi )
\\ \nonumber
&{\rm for}&
\qquad
n_{int} < n < n_{on} \; {\rm and} \; 0 < \xi < + \infty,
\end{eqnarray}
where
\begin{eqnarray}
\beta_1 &\equiv& -\frac{v}{2 D_0} +
\sqrt{ \left( \frac{v}{2 D_0} \right)^2
+ \frac{\Gamma_L}{\hbar D_0}\alpha_{RL}};
\\ \nonumber
\beta_2 &\equiv& \frac{v}{2 D_0} +
\sqrt{ \left( \frac{v}{2 D_0} \right)^2
+ \frac{\Gamma_L}{\hbar D_0}(\lambda + \alpha_{RL} + 1)}.
\end{eqnarray}

Continuity of the first derivative $dn/d \xi $ at $\xi = 0$
gives the condition
\begin{equation}
(\beta_1(v) + \beta_2(v))\frac{e \gamma u - E_w}{\lambda}
=
\beta_2(v) \frac{e \gamma u - E_w + E_e^F}{\lambda + \alpha_{RL} +1}.
\label{uv}
\end{equation}
This condition (\ref{uv}) implicitly determines the  dependence of
the front velocity $v$ on the applied voltage $u$.
Linearization of  this dependence
near the voltage
\begin{equation}
u_{co} = \frac{E_W}{e \gamma} +
\frac{1}{e \gamma}
\frac{\lambda E^F_e}
{1 + \alpha_{RL} + \sqrt{\alpha_{RL}} \sqrt{\lambda + \alpha_{RL} +
1}}
\label{u_co}
\end{equation}
that corresponds to $v=0$ leads to the explicit $v(u)-$dependence
(\ref{v(u)}).
The obtained dependence is equivalent to the linearization
of Eqn.(\ref{velocity}).

\end{multicols}
\vskip 1cm

\begin{figure}[tbp]
\setlength{\unitlength}{1cm}
\begin{center}
\begin{picture}(7,7)
\epsfxsize=7cm
\put(0,0){\epsffile{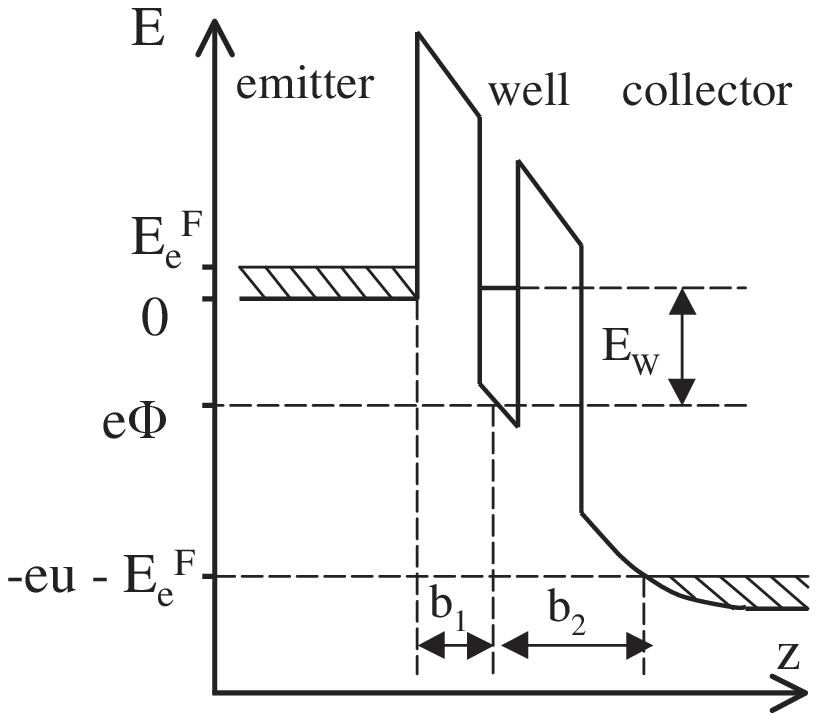}}
\end{picture}
\end{center}
\label{sketch}
\caption{
Schematic DBRT structure. Note that the applied voltage $u$
and the electron charge $e$ are chosen negative.
}
\end{figure}


\begin{figure}[tbp]
\setlength{\unitlength}{1cm}
\begin{center}
\begin{picture}(7,7)
\epsfxsize=7cm
\put(0,0){\epsffile{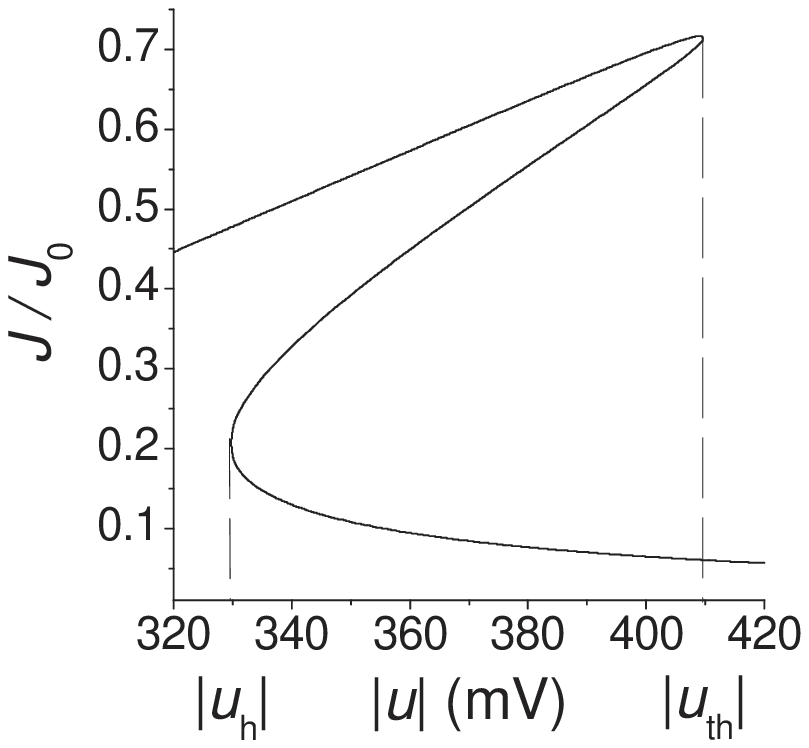}}
\end{picture}
\end{center}
\label{Fig.2}
\caption{
Stationary current-voltage characteristic of the DBRT
in the bistable regime. The current density
$J$ is normalized by $J_0 = e \Gamma_R \rho_0 E_e^F/ \hbar$.
This normalized current density is equivalent
to the normalized electron concentration in the well:
$J/J_0 \equiv n/( \rho_0 E_e^F)$.
The corresponding numerical values are
$J_0 \approx 7 \; kA/cm^2$ and
$ \rho_0 E_e^F \approx 3 \cdot 10^{11} \; cm^{-2}$
for the chosen parameters
$E_e^F = 10 \, {\rm meV}$, $b_1 = 10 \; {\rm nm}$, $b_2 = \; 50 {\rm nm}$,
$\Gamma_L = 0.5 \, {\rm meV}$, $\Gamma_R = 0.1 \, {\rm meV}$ and
$\Gamma_W = 2.0 \, {\rm meV}$.
}
\end{figure}

\newpage

\begin{figure}[tbp]
\setlength{\unitlength}{1cm}
\begin{center}
\begin{picture}(8,8)
\epsfxsize=8cm
\put(0,0){\epsffile{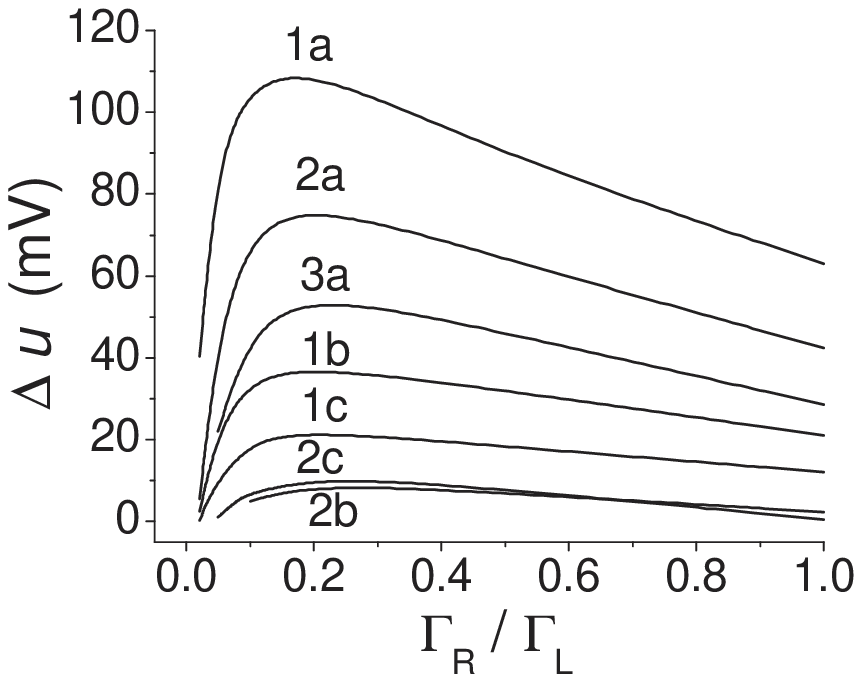}}
\end{picture}
\end{center}
\caption{
Bistability range as a function of $\Gamma_R/ \Gamma_L$. Parameters:
$E_e^F = 10 \, {\rm meV}, b_1 = 10 \; {\rm nm}$, $b_2 = 50 \; {\rm nm}$
and $\Gamma_W = 1 \, {\rm meV},\, 2 \, {\rm meV},\, 3 \, {\rm meV}$
for curves 1a, 2a, 3a, respectively;
$E_e^F = 5 \, {\rm meV},\, b_1 = 10 \, {\rm nm}$,\, $b_2 = \; 50 {\rm nm}$
and $\Gamma_W = 1 \, {\rm meV},\, 3 \, {\rm meV}$
for curves 1b, 2b, respectively;
$E_e^F = 5 \, {\rm meV},\, b_1 = 10 \; {\rm nm}$,\, $b_2 = \; 30 {\rm nm}$
and $\Gamma_W = 1 \, {\rm meV},\, 3 \, {\rm meV}$
for curves 1c and  2c, respectively.
}
\label{Fig.3}
\end{figure}


\begin{figure}[tbp]
\setlength{\unitlength}{1cm}
\begin{center}
\begin{picture}(7,7)
\epsfxsize=7cm
\put(-4,0){\epsffile{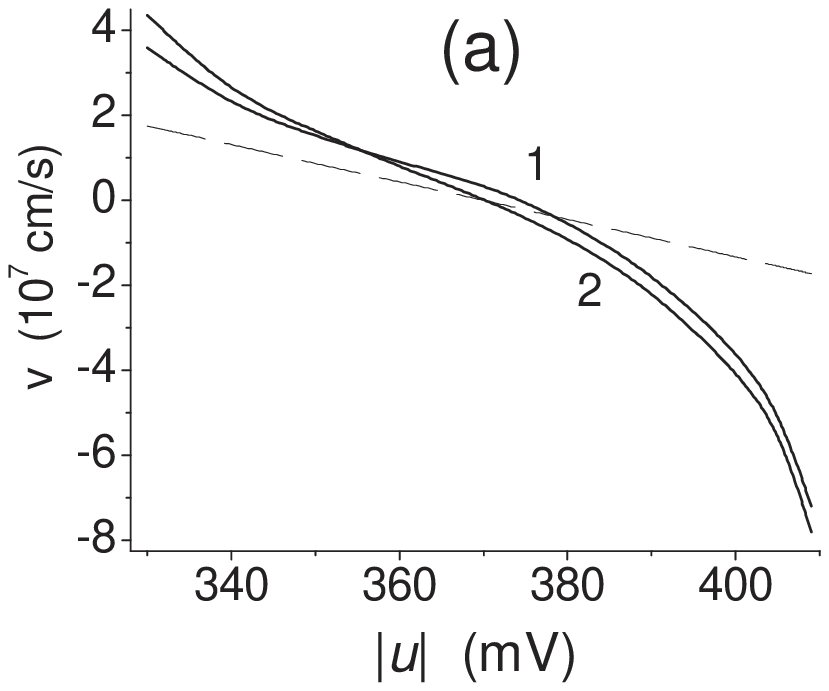}}
\epsfxsize=7cm
\put(4,0){\epsffile{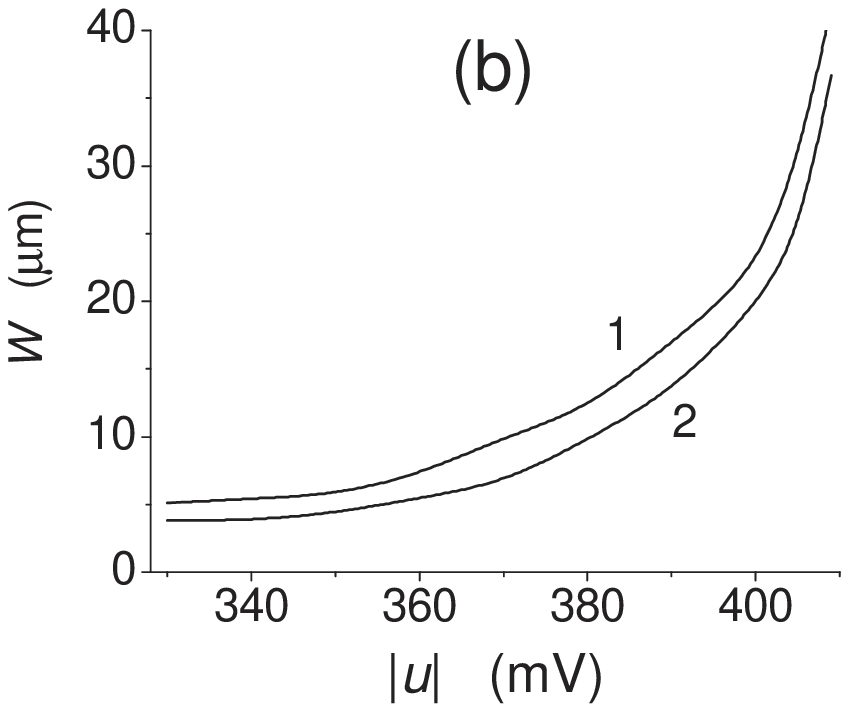}}
\end{picture}
\end{center}
\caption{
Front velocity (a) and front width (b)
as a function of the applied voltage $u$.
Curves 1 and 2 correspond to $T_W = 4 K$
and $T_W = 100 K$, respectively.
The dashed line in the panel (a) shows the prediction of the
analytical formula (\ref{v(u)2}) with $u_{co}$
approximated as $u_{co}=(u_{th}+u_{h})/2$.
Parameters as in Fig.2 :
$\Gamma_L = 0.5 \; {\rm meV}$,
$\Gamma_R = 0.1 \; {\rm meV}$,
$\mu = 10^5 \, {\rm cm^2/V \cdot s}$.
}
\label{Fig.4}
\end{figure}


\begin{thebibliography}{99}

\bibitem[*]{LEAVE}
On leave from Ioffe Physicotechnical
Institute, Politechnicheskaya 26, 194021, St.Petersburg,
Russia. Electronic mail: rodin@physik.tu-berlin.de

\bibitem{GOL87}
V. J. Goldmann, D. C. Tsui and J. E. Cunningham,
Phys. Rev. Lett. {\bf 58}, 1256 (1987).

\bibitem{MAR94} A. Martin, M. Lerch, P. Simmonds and
L. Eaves, Appl. Phys. Lett. {\bf 64}, 1248 (1994).

\bibitem{WAC95d}
A. Wacker and E. Sch{\"o}ll,
J. Appl. Phys. {\bf 78}, 7352 (1995).

\bibitem{GLA97}
B. Glavin, V. Kochelap and V. Mitin,
Phys. Rev. B {\bf 56} 13346 (1997).

\bibitem{MEL98}
D. Mel'nikov and A. Podlivaev,
Semiconductors {\bf32}, 206 (1998).

\bibitem{FEI98}
M. N. Feiginov and V. A. Volkov,
JETP Lett. {\bf 68}, 633 (1998)
[Pis'ma Zh. Eksp. Teor. Fiz. {\bf 68}, 628 (1998)].

\bibitem{MEI00}
M. Meixner, P. Rodin, E. Sch{\"o}ll and A. Wacker,
Eur. Phys. J B {\bf 13}, 157 (2000).

\bibitem{CHE00}
V. Cheianov, P. Rodin and E. Sch{\"o}ll,
Phys. Rev. B {\bf 62}, 9966 (2000).

\bibitem{SCH02}
E. Sch{\"o}ll, A. Amann, M. Rudolf and J. Unkelbach,
Physica B {\bf 314}, 113 (2002).

\bibitem{COMMENT_1}
The expression for $J_{ew}$ derived in Ref.\onlinecite{SCH02} agrees with
(\ref{current_density_1}) for $\Delta \gg kT_W, \Gamma_W$.
The expression for $J_{ew}$
derived in Ref.\onlinecite{MEI00} also includes an additional
logarithmic term
which gives a contribution of about 5 \% for the chosen
parameters. This term is skipped here for simplicity.

\bibitem{INK01} J. Inkoferer, G. Obermair and F. Claro,
Phys. Rev. B {\bf 64}, 201404 (2001).

\bibitem{DAT95}
S. Datta, {\it Electronic Transport in Mesoscopic Systems},
(Cambridge University Press, Cambridge 1995)

\bibitem{COMMENT_0}
The spacer layer in the emitter is not taken into
account here.

\bibitem{PRI92}
P. J. Price, {\it Electron Tunneling in
Semiconductors}, in {\it Handbook on Semiconductors},
(edited by P. T. Landsberg), Vol.1 (Elsevier, Amsterdam 1992).

\bibitem{GUE89}
P. Gu{\` e}ret, C. Rossel, E. Marclay, and H. Meier,
J. Appl. Phys. {\bf 66}, 278 (1989).


\bibitem{RID95} B. K. Ridley,
{Electrons and Phonons in Semiconductor Multilayers},
(Cambridge University Press, Cambridge 1995), p.23, eqn.(1.29).

\bibitem{KAR90}
V. Karpus,
Semicond. Sci. Technol. {\bf 5}, 691 (1990).

\bibitem{times}
Note that at $T = 4$~K
the momentum and energy relaxation times  in the quantum well are
determined by interface roughness scattering and
acoustic phonon scattering, respectively
~\cite{RID95,KAR90}.

\bibitem{COMMENT_GLA}
The monotonic dependence of the bistability range on the dimensionless
parameter $k = \Gamma_L/(\Gamma_L + \Gamma_R)$ has also been
reported in Ref.\onlinecite{GLA97} for the coherent tunneling model.

\bibitem{SCH00} E. Sch{\"o}ll,
{\em Nonlinear Spatio-Temporal Dynamics and Chaos in Semiconductors}
(Cambridge University Press, 2001).

\bibitem{WER99}
L.-E. Wernersson, M. Suhara, N. Carlsson, K. Furuya,
B. Gustafson, A. Litwin, L. Samuelson, and W. Seifert,
Appl. Phys. Lett. {\bf 74}, 311 (1999).

\bibitem{WER00}
L.-E. Wernersson, M. Borgstr{\"o}m, B. Gustafson, A. Gustafsson,
L. Jarlskog, J.-O. Malm, A. Litwin, L. Samuelson, and W. Seifert,
J. Crystal Growth {\bf 221}, 704 (2000).

\bibitem{MIT98}
I. Mitkov, K. Kladko and J. E. Pearson,
Phys.~Rev.~Lett.  {\bf 81}, 5453 (1998)

\bibitem{IMP}
A. M. Nechaev, B. F. Sinkevich,
Sov.~Phys.~Semicond., {\bf 18}, 218 (1984)
[Fiz.~Tekh.~Polupr. {\bf 18}, 350 (1984)].

\bibitem{ALE98}
A. Alekseev, S. Bose, P. Rodin and E. Sch{\"o}ll,
Phys.~Rev.~E {\bf 57}, 2640 (1998).

\bibitem{ZEE} S. M. Sze, {\it Modern Semiconductor Device Physics},
(Wiley, New York, 1998).

\bibitem{MEI98}
M. Meixner, P. Rodin and E. Sch{\"o}ll,
Phys.~Rev.~E  {\bf 58}, 2796 (1998).

\end{thebibliography}
\end{document}